\documentclass[10pt,preprintnumbers,aps,amssymb,nofootinbib,amsmath,superscriptaddress,notitlepage,prd]
{revtex4-2}
\usepackage{epsfig,epsf}
\usepackage{comment}
\usepackage{bm} 
\usepackage{color} 
\usepackage{slashed}
\usepackage{relsize}	
\usepackage{soul} 
\usepackage{hyperref}
\usepackage{tensor} 
\usepackage{yfonts} 
\newcommand{\beq}{\begin{equation}}
\newcommand{\beql}[1]{\begin{equation}\label{#1}}
\newcommand{\eeq}{\end{equation}}
\def\bal#1\gal{\begin{align}#1\end{align}}
\newcommand{\ball}[1]{\bal\label{#1}}
\newcommand{\eq}[1]{(\ref{#1})}

\renewcommand{\sec}[1]{Sec.~\ref{#1}}
\DeclareMathOperator{\sgn}{sgn}

\renewcommand{\b}[1]{{\bm #1}} 
\newcommand{\unit}[1]{\hat {{\bm #1}}} 

\DeclareMathOperator{\arctanh}{arctanh}
\begin{document}

\title{Scattering Amplitudes and Resonant Processes in QED with Chiral Chemical Potential and Chiral Magnetic Conductivity}

\author{Jonathan D. Kroth}

\affiliation{
Department of Physics and Astronomy, Iowa State University, Ames, Iowa, 50011, USA}

\author{Kirill Tuchin}

\affiliation{
Department of Physics and Astronomy, Iowa State University, Ames, Iowa, 50011, USA}

\date{\today}


\begin{abstract}

The QED scattering amplitude in a chiral medium characterized by a constant chiral chemical potential $\mu_5$ and chiral magnetic conductivity $b_0$ is analyzed. We show the emergence of the resonant behavior in $1\to 2$, $2\to 2$, and $2\to 3$ processes. We compute the rates of paradigm $1\to 2$ processes that determine the widths of quasi-stationary fermion and photon states in the medium. We elucidate the origin of these resonances, the conditions of their emergence, and the physical principles of their regularization.

\end{abstract}

\maketitle


\section{Introduction}\label{sec:a}

A medium containing chiral fermions may be chiraly imbalanced. The chiral imbalance can be expressed in terms of the $P$ and $CP$-odd terms in the Lagrangian. In quark-gluon plasma, the net chirality is reflected in the chiral chemical potential $\mu_5$ induced by the sphaleron transitions. In Weyl semimetals, the chirality is manifested as a displacement $\b \Delta$ of the Weyl nodes in momentum space. Many phenomenological models describe the chirality by introducing  a term into Lagrangians that couples the four vector $(\mu_5,  \b \Delta)$ to the axial vector current $j^{\mu 5}=\bar \psi \gamma^\mu\gamma^5\psi$. 

In the presence of a chiral imbalance, two new currents are induced in the medium due to the chiral anomaly \cite{Adler:1969gk,Bell:1969ts}. These currents are the chiral magnetic current $\b j = b_0\b B$ \cite{Kharzeev:2004ey,Kharzeev:2007tn,Fukushima:2008xe,Kharzeev:2009fn,Kharzeev:2007jp} and the anomalous Hall current $\b j = \b b\times \b E$, where $b_0=c_A\mu_5$ is called the chiral magnetic conductivity and $\b b= c_A\b \Delta/2$. A medium with the $P$ and $CP$-odd responses defined by the chiral magnetic and anomalous Hall currents can be described employing constitutive relations. These relations are linear when $\mu_5$ and $\b \Delta$ are constant, and they become isotropic when  $\b \Delta$ vanishes. From the field-theoretic perspective, the anomalous currents can be represented as the Chern-Simons term $\tilde F F$ in the Lagrangian, where $\tilde F$ is  the dual electromagnetic field tensor.

To obtain an effective Lagrangian for electrodynamics with chiral imbalance, perform a chiral transformation $\psi\to e^{i\theta\gamma^5}\psi$ of the fermionic functional integral. This transformation produces the following new terms in the Lagrangian \cite{Fujikawa:2004cx}:
\ball{a1}
 \Delta \mathcal L= \theta(x)\left\{ \partial_\mu j^{\mu 5}-c_A\left(-\frac{1}{4}\tilde F^{\mu\nu}F_{\mu\nu}\right)\right\}\,,
 \gal
where 
\ball{a3}
c_A=\frac{1}{2\pi^2}\sum_ae_a^2
\gal
is the chiral anomaly coefficient, with $e_a$ being the electric charge of fermion $a$ and sum running over all fermion species. The expression in the curly brackets in \eq{a1} is the total derivative, and the Lagrangian \eq{a1} can be cast in the following form:
\ball{a5}
 \Delta \mathcal L=   \partial_\mu\theta \left\{ -j^{\mu 5}  +    \frac{c_A}{2}\epsilon^{\mu\nu\lambda\rho}A_\nu\partial_\lambda A_\rho \right\}\,.
 \gal
Therefore, the equations of motion depend not on $\theta$ directly, but on its derivative $\partial_\mu \theta$. It is a simple exercise to verify that $b_\mu=c_A\partial_\mu \theta$, with $b_0$ and $\b b$ being its temporary and spacial components respectively.  

With the addition of $\Delta\mathcal{L}$ to the Lagrangian of QED, we obtain an effective theory that is extensively used in applications.
The chiral magnetic effect, its theory, and its applications are discussed in \cite{babaev2023chiral}. A review focusing on the phenomenology of relativistic heavy-ion collisions is presented in \cite{Kharzeev:2015znc,Kharzeev:2024zzm}, a review on topological materials is presented in \cite{Sekine:2020ixs,Ong:2020ffe} while reference \cite{Sikivie:2020zpn} delves into axion phenomenology. 

The goal of this article is to explore the role of the chiral imbalance played in scattering processes using the effective theory presented above. In particular, we  argue that scattering amplitudes often exhibit resonant behavior in certain chiral modes. A typical example is the chiral (or vacuum) Cherenkov radiation whose rate is characteristically proportional to the delta function of the emitted transverse momentum. We will elucidate the origin of these resonances, the conditions of their emergence and their regularization procedure. 

Most applications we have in mind, pertain to high energy nuclear and particle physics where it is usually assumed that chirality distribution is uniform, i.e.\ $\b\Delta=0$. We adopt this approximation as it allows us to present our arguments in the clearest form. However, all conclusions can be readily generalized to any $\b \Delta$. Furthermore, we work in the high energy approximation for the same reason.

The paper is structured as follows. \sec{sec:f} and \sec{sec:b} review the solutions to the Maxwell equations in a medium with a chiral magnetic current and the Dirac equation at finite chiral chemical potential. In each case, we derive the chiral plane waves and propagators. The detailed derivations are presented in Appendices \ref{sec:AppA} and \ref{sec:AppE}. \sec{sec:12} deals with the chiral Cherenkov radiation and its cross-channel, pair production. We compute the production rates, show the emergence  of the resonant behavior, and study its dependence on $\mu_5$ and $b_0$. \sec{sec:j} investigates the resonances in $2\to 2$ scattering, and section \sec{sec:k} does the same for bremsstrahlung. A summary is presented in \sec{sec:s}.

We use natural units $c=\hbar=1$ and convention of \cite{Peskin:1995ev} for Dirac matrices. 

\section{Solution to Maxwell equations in the presence of  chiral magnetic current}\label{sec:f}

The chiral magnetic current is a component of the medium's linear response to an external electromagnetic field. Qiu, Cao and Huang  incorporate it into the  constitutive relationship between the displacement and electric fields  \cite{Qiu:2016hzd}. The primary concern in this section is the electromagnetic field in a medium exhibiting such anomalous constitutive relationship. We ignore all medium effects and sources apart from the chiral magnetic current.

\subsection{Maxwell equations with chiral magnetic current}

Maxwell's equations, obtained from \eq{a5} read:
\begin{subequations}\label{f1}
\bal
&\b \nabla \times \b B = \dot{\b E} +b_0 \b B  \,,\label{f2}\\
&\b \nabla\cdot \b E= 0\,,\label{f3}\\
&\b \nabla \times \b E =-\dot{\b B}\,,\label{f4}\\
&\b \nabla\cdot \b B=0\,.\label{f5}
\gal 
\end{subequations}
In the radiation gauge $A^0=0$, $\b \nabla\cdot A=0$ \eq{f2} yields the modified wave equation for the vector potential:
\ball{f7} 
-\ddot{\b A}+\nabla^2\b A+b_0\b\nabla\times \b A=0\,.
\gal


The positive energy solutions of \eq{f7} have the form 
\ball{f9}
\b A_{\b p, \lambda}(x)= \frac{1}{\sqrt{2\omega_{\b p, \lambda}}}\b\epsilon_{\b p, \lambda} e^{-ip\cdot x}\,,
\gal 
where $p^\mu= (\omega_{\b p, \lambda}, \b p)$, and $\b\epsilon_{\b p, \lambda}$ is the polarization vector corresponding to  the polarization $\lambda$. Substituting \eq{f9} into \eq{f7} we readily find 
\ball{f11}
\left( \omega_{\b p, \lambda}^2-\b p^2\right)\b\epsilon_{\b p, \lambda}+b_0i\b p\times \b\epsilon_{\b p, \lambda}=0\,.
\gal
It order to diagonalize this equation, we must choose the polarization vectors to represent the right  and left circular polarizations corresponding to $\lambda=\pm 1$. This is because such vectors obey the identity:
\ball{f13}
i \unit p\times \b\epsilon_{\b p, \lambda}= \lambda\b\epsilon_{\b p, \lambda}\,.
\gal
Using \eq{f13} in \eq{f11} yields the dispersion relation:
\ball{f15}
\omega_{\b p, \lambda}=  \sqrt{\b p^2-\lambda b_0|\b p|}\,.
\gal

The negative energy solutions read:
\ball{f17} 
\b A'_{\b p, \lambda}(x)= \frac{1}{\sqrt{2\omega'_{\b p, \lambda}}}\b\epsilon'_{\b p, \lambda} e^{ip\cdot x}\,,
\gal
The dispersion of the negative energy solutions is found by plugging \eq{f17} into \eq{f7}, which yields
\ball{f19}
\left( {\omega'}^2_{\b p, \lambda}-\b p^2\right)\b\epsilon'_{\b p, \lambda}-b_0i\b p\times \b\epsilon'_{\b p, \lambda}=0\,.
\gal
Choosing $\b\epsilon'_{\b p, \lambda}=\b\epsilon^*_{\b p, \lambda}$ and using  the complex conjugate of \eq{f13} we find that $\omega'_{\b p, \lambda} = \omega_{\b p, \lambda}$. As a result, 
\ball{f23}
\b A'_{\b p, \lambda}(x)= \b A^*_{\b p, \lambda}(x)\,.
\gal
Therefore, the total vector potential, which is the sum of \eq{f9} and \eq{f17} is real, as required.

The quantized electromagnetic field reads:
\ball{f27}
\b A(x)= \sum_\lambda \int \frac{d^3p}{(2\pi)^3}\frac{1}{\sqrt{2\omega_{\b p, \lambda}}}\left[ \b\epsilon_{\b p, \lambda} e^{-ip\cdot x}a_{\b p, \lambda} + \b\epsilon^*_{\b p, \lambda} e^{ip\cdot x}a^\dagger_{\b p, \lambda} \right]\,,
\gal
with the commutation relations given by
\ball{f29}
[a_{\b p, \lambda},a^\dagger_{\b p', \lambda'}]= (2\pi)^3\delta(\b p-\b p')\delta_{\lambda,\lambda'}\,.
\gal

The energy and the Hamiltonian of electromagnetic field is discussed in Appendix~\ref{sec:AppA}. The quantization of the Maxwell theory with the Chern-Simons term was discussed in \cite{Colladay:1998fq,Adam:2001ma}.

The photon's Feynman propagator in a covariant gauge for an arbitrary $b_\mu$ is given by \cite{Carroll:1989vb,Lehnert:2004hq}:
\ball{f60}
\tilde {\mathcal{D}}_F^{\mu\nu}(k)= \frac{-i k^2 g^{\mu\nu}+ \epsilon^{\mu\nu\rho\sigma}b_\rho k_\sigma-ib^\mu b^\nu}{k^4+k^2b^2-(b\cdot k)^2+i\epsilon}\,.
\gal
In Appendix~\ref{sec:AppA} we present a detailed derivation of the photon propagator in the Coulomb gauge, starting from the corresponding correlation function. 

Let us now examine the poles of the photon propagator \eq{f60}. At $\b b=0$, the propagator has poles at  
\ball{f66}
(k^0)^2=\b k^2+\lambda b_0 |\b k|\,.
\gal 
In particular, in static limit $k^0=0$, besides the familiar Coulomb pole at $|\b k|=0$, there is a pole at $|\b k| =  b_0$ indicating the instability of the magnetic field in the chiral matter \cite{Carroll:1989vb,Joyce:1997uy,Adam:2001ma}. This instability is not the subject of this study, although it is related to the resonances we discuss in the subsequent sections. We elaborate on this connection in \sec{sec:s}. 

At $b_0=0$, the denominator of \eq{f60} vanishes when
\ball{f67}
\b k^4+ \b k^2\b b^2-(\b b\cdot \b k)^2=0\,.
\gal 
This implies the presence of only one pole at $|\b k|=0$, in contrast to the previous case. Consequently, the magnetic field is stable in this situation. This is further discussed in \sec{sec:s}.

\section{Solution to Dirac equation at finite chiral chemical potential}\label{sec:b}

The Dirac equation at finite $\mu_5$ reads:
\ball{b1}
(i\slashed \partial - \mu_5\gamma^5\gamma^0 -m)\psi=0\,.
\gal 
Its solutions were obtained and discussed in \cite{Colladay:1996iz,Kostelecky:2002ue,Kharzeev:2007jp,Sukhachov:2018uuz,Sheng:2017lfu}. More general cases involving finite $\b \Delta$ were explored in \cite{Colladay:1996iz,Kroth:2026kgm}.

The solutions of \eq{b1} are derived in Appendix~\ref{sec:AppE}. The positive and negative energy spinors read:
\ball{b24}
u_{\b p, \sigma}= 
\left(\begin{array}{c} 
\sqrt{E_{\b p, \sigma}-\sigma|\b p|+\mu_5} \xi _{\b p, \sigma}\\ 
\sqrt{E_{\b p, \sigma}+\sigma|\b p|-\mu_5} \xi _{\b p, \sigma}
\end{array}\right)\,,
\gal
with the dispersion
\ball{b16-1}
E_{\b p,\sigma}= \sqrt{(\sigma|\b p|-\mu_5)^2+m^2}\,,
\gal
and
\ball{c24}
v_{\b p, \sigma}= 
\left(\begin{array}{c} 
\sqrt{E'_{\b p, \sigma}-\sigma|\b p|-\mu_5} \xi _{\b p, \sigma}\\ 
-\sqrt{E'_{\b p, \sigma}+\sigma|\b p|+\mu_5} \xi _{\b p, \sigma}
\end{array}\right)\,,
\gal
with the dispersion $E'_{\b p,\sigma}=E_{\b p,-\sigma}$ respectively. 

Employing the identities
\ball{g1}
\left(E_{\b p, \sigma}+m\mp |\b p|\sigma\pm \mu_5\right)^2= 2(E_{\b p, \sigma}+m)\left(E_{\b p, \sigma}\mp|\b p|\sigma\pm\mu_5\right)\,,
\gal
\eq{b24} can be cast in a different form:
\ball{g3}
u_{\b p, \sigma}= 
\frac{1}{\sqrt{2(E_{\b p, \sigma}+m)}}
\left(\begin{array}{c} 
\left(E_{\b p, \sigma}-\sigma|\b p|+m+\mu_5\right) \xi _{\b p, \sigma}\\ 
\left(E_{\b p, \sigma}+\sigma|\b p|+m-\mu_5\right) \xi _{\b p, \sigma}
\end{array}\right)\,.
\gal
Similarly, the anti-particle spinors can be represented as:
\ball{g11}
v_{\b p, \sigma}= 
\frac{1}{\sqrt{2(E'_{\b p, \sigma}+m)}}
\left(\begin{array}{c} 
\left(E'_{\b p, \sigma}-\sigma|\b p|+m-\mu_5\right) \xi _{\b p, \sigma}\\ 
-\left(E'_{\b p, \sigma}+\sigma|\b p|+m+\mu_5\right) \xi _{\b p, \sigma}
\end{array}\right)\,.
\gal
The spinors in the form \eq{g3} and \eq{g11} are convenient for the high-energy expansion.

A derivation of the fermion propagator is presented in Appendix~\ref{sec:AppE}. In momentum space it reads 
\ball{d12}
\tilde G(p)&=\frac{i\left(\slashed p -  \gamma^5\gamma^0\mu_5 +m\right)\left(p^2-\mu_5^2-m^2-2\mu_5\b \Sigma \cdot \b p\right)}{\left(p^2-\mu_5^2-m^2\right)^2-4\mu_5^2\b p^2}\,.
\gal
 A more general expression can be found in \cite{Miransky:2015ava}.
$\tilde G(p)$ has four simple poles at $p^0= \pm E_{\b p,\sigma}$, where $E_{\b p,\sigma}$ are given by \eq{b16-1}. Adding the $+i\epsilon$ prescription to the denominator transforms \eq{d12} into the Feynman propagator.

At high energies, where $|\b p|\gg \mu_5, m$, the dispersion relation \eq{b16-1} reads
\ball{g5}
E_{\b p, \sigma}\approx |\b p|-\sigma\mu_5\,.
\gal
Substituting this into \eq{g3} and expanding  yields the high-energy approximation for the particle spinor:
\ball{g7}
u_{\b p, \sigma}\approx  
\frac{1}{\sqrt{2p_z}}
\left(\begin{array}{c} 
\left[|\b p|(1-\sigma)+\frac{1}{2}\mu_5(1-\sigma)+\frac{1}{2}m(1+\sigma)\right] \xi _{\b p, \sigma}\\[1mm]
\left[|\b p|(1+\sigma)-\frac{1}{2}\mu_5(1+\sigma)+\frac{1}{2}m(1-\sigma)\right] \xi _{\b p, \sigma}
\end{array}\right)\,.
\gal
This formula can be  directly derived from \eq{b24}, but it requires keeping the next term, proportional to $m^2$, in the expansion of \eq{g5}. 

By the same token, expanding \eq{g11} yields the high-energy approximation of the anti-particle spinor:
\ball{g13}
v_{\b p, \sigma}\approx  
\frac{1}{\sqrt{2p_z}}
\left(\begin{array}{c} 
\left[|\b p|(1-\sigma)-\frac{1}{2}\mu_5(1-\sigma)+\frac{1}{2}m(1+\sigma)\right] \xi _{\b p, \sigma}\\[1mm]
-\left[|\b p|(1+\sigma)+\frac{1}{2}\mu_5(1+\sigma)+\frac{1}{2}m(1-\sigma)\right] \xi _{\b p, \sigma}
\end{array}\right)\,.
\gal

Finally, assuming that $|\b p|\approx p_z$, the helicity eigenstates read:
\ball{g19}
\xi_{\b p,+}\approx  \left(\begin{array}{c} 1 \\ \frac{p_x+ip_y}{2p_z}  \end{array}\right)\,,\qquad
\xi_{\b p,-}\approx \left(\begin{array}{c} -\frac{p_x-ip_y}{2p_z} \\ 1  \end{array}\right)\,.
\gal

\section{$1\to 2$ scattering}\label{sec:12}

This section calculates the rates of two $1\to 2$ processes, $q\to q\gamma$ and $\gamma\to q\bar q$, in the presence of both the chiral magnetic current \emph{and} the chiral chemical potential $\mu_5$. Here, $q$ and $\bar q$ represent fermions and antifermions, respectively. The latter process was first discussed in \cite{Lehnert:2004hq,Lehnert:2004be} in the context of Lorentz symmetry-violating theories, assuming $\mu_5=0$. Dubbed the vacuum Cherenkov radiation, it is regarded as a signature of such violation \cite{Carroll:1989vb,Lehnert:2004hq,Lehnert:2004be,Klinkhamer:2004hg,Mattingly:2005re,Kostelecky:2002ue,Jacobson:2005bg,Altschul:2006zz,Altschul:2007kr,Nascimento:2007rb}. 

The full QFT calculation at $\mu_5=0$ was performed by one of us in \cite{Tuchin:2018sqe}. In that work, we  proposed these processes as a manifestation of the $P$ and $CP$ violation in chiral media. In this context, they are termed the chiral Cherenkov radiation and pair production. Extensive studies of these processes have been conducted in \cite{Tuchin:2018sqe,Huang:2018hgk,Tuchin:2018mte,Hansen:2020irw,Hansen:2022nbs,Hansen:2023wzp,Hansen:2024kvc,Hansen:2024xdg}. Additionally, the cascade induced by these novel channels has been studied in \cite{Hansen:2025gzt}. Furthermore, the non-Abelian version of the chiral Cherenkov radiation has been explored in \cite{Hansen:2024rlj}, based on the non-Abelian generalization of the chiral magnetic effect, as discussed in \cite{Akamatsu:2013pjd,Duari:2025kar}. The chiral processes induced by anomalous Hall current were discussed in \cite{Huang:2018hgk,Tuchin:2018mte,Hansen:2024kvc,Tuchin:2025stl}. Lastly, the classical version of the chiral Cherenkov radiation has been studied in \cite{Hansen:2020irw,Barredo-Alamilla:2023xdt,vonDossow:2025fbb}.

\subsection{Chiral Cherenkov radiation}\label{sec:h}

Consider the reaction $q(\b p,\sigma)\to q(\b p',\sigma')+ \gamma(k,\lambda)$ in the frame where the incident quark moves along the $z$-axis:\footnote{In this section, the prime denotes the quantum numbers of the final fermion, unlike in \sec{sec:b} where it labels the anti-particle states. }
\begin{subequations}\label{h5}
\bal
 p&=(E_{\b p,\sigma},0,0,p_z)\,,\\ 
 k&=(\omega_{\b k,\lambda},k_\bot,0,xp_z)\,,\\
  p'&=(E'_{\b p',\sigma'},-k_\bot,0,(1-x)p_z)\,.
\gal
\end{subequations}
 The rate of the photon radiation is given by 
\ball{h9}
d\dot w = \frac{1}{2}\frac{1}{(2\pi)^3}\sum_{\lambda,\sigma,\sigma'}\frac{1}{8E^3x(1-x)}2\pi \delta\left(E_{\b p,\sigma}-E'_{\b p',\sigma'}-\omega_{\b k, \lambda}\right)|i\mathcal M|^2d^2k_\bot dx p_z\,,
\gal
where the matrix element is
\ball{h11}
i\mathcal{M}= -ie\bar u_{\b p',\sigma'}\slashed \epsilon^*_{\b k,\lambda} u_{\b p, \sigma}\,.
\gal

At high energies, $p_z\gg m\sim k_\bot\sim  \mu_5\gg b_0$. In other words, let  $\varepsilon$ be a small bookkeeping parameter. Then $p_z\sim \mathcal{O}(\varepsilon^0)$, $m\sim k_\bot\sim \mu_5\sim \mathcal{O}(\varepsilon^1)$, and $b_0\sim\mathcal{O}(\varepsilon^2)$. Expanding the time components of the three four-vectors in \eq{h5} up to and including the order $\varepsilon^2$ yields the following approximations:
\begin{subequations}
\bal
E_{\b p,\sigma}&\approx p_z\left( 1-\frac{\sigma\mu_5}{p_z}+\frac{m^2}{2p_z^2}\right)\,,\label{h13}\\
E'_{\b p',\sigma'}&\approx p_z'\left( 1-\frac{\sigma'\mu_5}{p'_z}+\frac{m^2+k_\bot^2}{2{p'_z}^2}\right)\,,\label{h13a}\\
\omega_{\b k, \lambda}&\approx k_z\left( 1-\frac{\lambda b_0}{2k_z}+\frac{k_\bot^2}{2k_z^2}\right)\,.\label{h13b}
\gal
\end{subequations}


The scattering amplitude $i\mathcal{M}$ is proportional to the matrix element $\xi_{\b p',\sigma'}^\dagger \b \sigma\cdot \b \epsilon^*_{\b k,\lambda} \xi_{\b p, \sigma}$. Using the expressions for 
two-component spinors $\xi$ given by \eq{g19}, we find:
\ball{h21}
\xi_{\b p',\sigma'}^\dagger \b \sigma \xi_{\b p, \sigma}
=\left[ \frac{p_x'}{2p_z'}(\sigma\unit x+i\unit y)+\sigma \unit z\right]\delta_{\sigma,\sigma'}
+\left[ \unit x+i\sigma\unit y -\frac{p_x'}{2p_z'}\unit z\right]\delta_{\sigma,-\sigma'}\,.
\gal
The photon polarization vector is given by $\epsilon_{\b k,\lambda} = (0, \b\epsilon_{\b k,\lambda})$ with 
\ball{h23}
\b\epsilon_{\b k,\lambda} = \frac{1}{\sqrt{2}}\left( 1, i\lambda, -\frac{k_\bot}{k_z}\right)\,.
\gal
It follows that 
\ball{h25}
\xi_{\b p',\sigma'}^\dagger \b \sigma\cdot \b \epsilon^*_{\b k,\lambda} \xi_{\b p, \sigma}
=-\frac{k_\bot}{\sqrt{2}p_z}\frac{\lambda x+\sigma(2-x)}{2x(1-x)}
\delta_{\sigma,\sigma'}
+\frac{1}{\sqrt{2}}(1+\sigma\lambda)\delta_{\sigma,-\sigma'}\,.
\gal

Substituting \eq{g7} into \eq{h11} and noticing that the helicity preserving term in \eq{h25} contains an extra factor of $1/p_z$, yields, after some algebra,
\ball{h19}
i\mathcal{M}&= \frac{ie}{2p_z\sqrt{1-x}}\xi_{\b p',\sigma'}^\dagger \b \sigma\cdot \b \epsilon^*_{\b k,\lambda} \xi_{\b p, \sigma}
\left[ -4p_z^2(1-x)\sigma \delta_{\sigma,\sigma'}- 2mp_z x\sigma\delta_{\sigma,-\sigma'} \right]\,,\nonumber\\
&= \frac{ie}{\sqrt{2(1-x)}}\left[ k_\bot \frac{\lambda x \sigma+2-x}{x}\delta_{\sigma,\sigma'}-mx(\sigma+\lambda)\delta_{\sigma,-\sigma'}\right]\,.
\gal
The sub-eikonal term proportional to $\mu_5/p_z$ has been neglected. Remarkably, \eq{h19} exhibits no dependence on either $b_0$ or $\mu_5$.

The argument of the delta-function in \eq{h9} expresses energy conservation:
\ball{h15}
E_{\b p,\sigma}-E'_{\b p',\sigma'}-\omega_{\b k, \lambda}= -\frac{k_\bot^2+m^2x^2}{2p_z x(1-x)}+\frac{1}{2}\lambda b_0-(\sigma-\sigma')\mu_5\,.
\gal
It vanishes only for certain values of the quantum numbers of the incident and produced particles. To simplify the notation, assume that $b_0$ and $\mu_5$ are positive (they must have the same sign). There are two possible channels allowed by energy conservation.  The first one occurs when the quark helicity does not change: $\sigma=\sigma'$ and the photon is right-hand polarized: $\lambda=+1$. In this case the differential rate reads:
\ball{h31}
\frac{d\dot w(\sigma=\sigma')}{d^2k_\bot dx}=\frac{e^2}{4\pi}\frac{k_\bot^2}{4\pi E}\frac{x^2-2x+2}{(1-x)x^2}
\delta\left(k_\bot^2-K_0^2\right)\theta\left(K_0^2\right)\,,
\gal
where 
\ball{h32}
K_0^2= \lambda b_0 E x(1-x)-m^2x^2\,.
\gal
The $m^2$ term in \eq{h32} must be retained because $b_0E\sim m^2\sim \mathcal{O}(\varepsilon^2)$.

In the second case, the quark spin flips as follows: $\sigma=-1$, $\sigma'=1$. The last term in \eq{h25} indicates that the only possible photon polarization is $\lambda=-1$. The differential rate reads:
\ball{h35}
\frac{d\dot w(\sigma=-\sigma'=-1)}{d^2k_\bot dx}=\frac{e^2}{4\pi}\frac{m^2}{4\pi E}\frac{x^2}{(1-x)}
\delta(k_\bot^2-K^2)\theta\left(K^2\right)\,,
\gal
where 
\ball{h36}
K^2= (4\mu_5+\lambda b_0) E x(1-x)-m^2x^2\,.
\gal
 The opposite scenario with $\sigma=1$, $\sigma'=-1$ is not possible because \eq{h15} cannot be satisfied given that $b_0\ll \mu_5$.

The energy spectrum of the rate is obtained by  adding \eq{h31} and \eq{h35} and integrating over the photon transverse momentum, which yields: 
\ball{h39}
\frac{d\dot w}{dx}&= \frac{e^2}{4\pi}\frac{1}{4E}\left[K_0^2\frac{x^2-2x+2}{(1-x)x^2}\theta\left(K_0^2\right)+\frac{m^2x^2}{1-x}\theta\left(K^2\right)\right]\,.
\gal 
The total rate has the familiar QED infrared divergence at $x\to 0$. By parameterizing it with a cutoff $\omega_c$, we obtain the rate, with the logarithmic accuracy, 
\ball{h40}
\dot w=\frac{e^2}{4\pi}\frac{\lambda b_0}{2}\log \left[\frac{\lambda b_0 E^2}{(\lambda b_0 E+m^2)\omega_c}\right]\delta_{\lambda,\sgn b_0}\,.
\gal
The total radiated energy, which can be computed by integrating over $x$ the product of \eq{h39} with $xE$,  is independent of $\omega_c$. 

In the chiral limit, only the helicity-preserving channel contributes to the rate. In a model with  $\mu_5=0$ and finite $b_0$, $K$ coincides with  $K_0$, and \eq{h39} simplifies: 
\ball{h41}
\frac{d\dot w}{dx}\Big|_{\mu_5=0}&=\frac{e^2}{4\pi}\frac{1}{4E}\left[ \lambda b_0 E\frac{(1-x)^2+1}{x}-2m^2\right]\theta\left(K_0^2\right)\,.
\gal
This result was previously derived in \cite{Tuchin:2018sqe}.

\subsection{Pair production}\label{sec:m}

The cross channel of the chiral Cherenkov radiation is the pair production. Assuming that the incident photon moves in the $z$-direction, and $x$ is the quark's longitudinal momentum momentum fraction, the differential quark production rate is given by: 
\ball{m0}
d\dot w = \frac{1}{2}\frac{1}{(2\pi)^2}\sum_{\lambda,\sigma,\sigma'}\frac{k_z}{8\omega^3x(1-x)} \delta\left(E_{\b p,\sigma}+E'_{\b p',\sigma'}-\omega_{\b k, \lambda}\right)|i\mathcal M|^2d^2p_\bot dx \,,
\gal
where now $x=p_z/k_z$ is the fraction of the initial momentum carried by the quark. The amplitude is given by:
\ball{m1}
i\mathcal{M}&= \frac{ie}{\sqrt{2x(1-x)}}\left\{ 
(2x-1-\lambda \sigma)p_\bot\delta_{\sigma',-\sigma}+
m(\sigma+\lambda)\delta_{\sigma',\sigma}
\right\}\,.
\gal
The energy conservation is expressed by the equation:
\ball{m3}
E_{\b p,\sigma}+E'_{\b p',\sigma'}-\omega_{\b k, \lambda}\approx -\mu_5(\sigma+\sigma')+\frac{1}{2}\lambda b_0+\frac{m^2+p_\bot^2}{2x(1-x)\omega}\,,
\gal
were we used \eq{h13}--\eq{h13b}.

In the channel where both quark and anti-quark have the same helicity $\sigma=\sigma'$, the quark production rate reads:
\ball{m5}
\frac{d\dot w(\sigma=\sigma')}{d^2p_\bot dx}= \frac{e^2}{4\pi}\frac{m^2}{4\pi \omega}\frac{1}{x(1-x)}\delta_{\lambda,\sigma}\delta(p_\bot^2-P^2)\theta(P^2)\,,
\gal
where 
\ball{m6}
P^2=(4\mu_5\sigma- \lambda b_0)x(1-x)\omega-m^2\,.
\gal
Evidently, $P^2$ is positive only if $4\mu_5\sigma- \lambda b_0>0$. In particular, at finite $\mu_5$, $P^2$ is positive only for the helicity $\sigma = \sgn\mu_5 $. 

In the channel where the quark and anti-quark have opposite helicities $\sigma=-\sigma'$, we obtain:
\ball{m7}
\frac{d\dot w(\sigma\neq \sigma')}{d^2p_\bot dx}= \frac{e^2}{4\pi}\frac{p_\bot^2}{4\pi\omega}\frac{(1-x)^2+x^2}{ x(1-x)}\delta(p_\bot^2-P_0^2)\theta(P_0^2)\,,
\gal
where
\ball{m8}
P_0^2=- \lambda b_0 x(1-x)\omega-m^2\,.
\gal
In this case only the photon polarization $\lambda= -\sgn b_0$ contributes regardless of the quark and anti-quark helicities. 

Adding \eq{m5} and \eq{m8} and integrating over $p_\bot$ and $x$ yields the total rate:
\ball{m11}
\dot w = \frac{e^2}{4\pi} \frac{1}{6\omega}\left\{\left[
 \sqrt{|\lambda b_0|\omega(|\lambda b_0|\omega-4m^2)}
-6m^2\arctanh\sqrt{1-\frac{4m^2}{|b_0\lambda|\omega}}+
2m^2\sqrt{1-\frac{4m^2}{|b_0\lambda|\omega}}
\right]\delta_{\lambda,-\sgn b_0}
\right.  \nonumber\\
\left. +6m^2\arctanh\sqrt{1-\frac{4m^2}{(4\mu_5 \sigma-b_0\lambda)\omega}}\right\}\,.
\gal
In the limit of very high energies $\omega\gg 4m^2/b_0$, the total rate \eq{m11} is  independent of energy and $\mu_5$: 
\ball{m13}
\dot w = \frac{e^2}{4\pi} \frac{|\lambda b_0|}{6}\delta_{\lambda,-\sgn b_0}\,.
\gal

In a model with  $\mu_5=0$ and $b_0\neq 0$, $P=P_0$ and one obtains after integrating over the quark's transverse momentum \cite{Tuchin:2018sqe}:
\ball{m15}
\frac{d\dot w}{dx}\Big|_{\mu_5=0}= \frac{e^2}{4\pi}\frac{|\lambda b_0|}{4}\left[ (1-x)^2+x^2+\frac{2m^2}{|\lambda b_0|\omega}\right]\theta\left(P_0^2\right)\,.
\gal

We only considered $1\to 2$  processes, however $2\to 1$ processes are also allowed. The corresponding cross sections can be obtained by crossing one of the final particles to the initial state. However, the phase space of such scattering is quite small, as it is proportional to a $\delta$-functions expressing energy conservation.

\section{$2\to 2$ scattering}\label{sec:j}

Consider a generic $2\to 2$ scattering: $p_1p_2\to p_3p_4$. The anomalous terms  give a small contribution unless the scattering amplitude has a resonance, which occurs when the denominator of a propagator vanishes. In this section, we will study the tree-level amplitudes at high energies and derive the conditions that allow the emergence of resonances.

Choose the center-of-mass frame:
\ball{j3}
p_1=(E_1, k\unit z)\,, \qquad p_2=(E_2,-k\unit z)\,,\qquad p_3=(E_3, k'\unit n)\,,\qquad p_4=(E_4, -k'\unit n)\,.
\gal 
where $a=1,\ldots,4$, 
\ball{j5}
E_a=\sqrt{\b p_a^2-2\Lambda_a |\b p_a|+M^2_a}\,.
\gal
We introduced notations:
\ball{j7}
\Lambda_a= 
\left\{\begin{array}{ll}
\sigma_a\mu_5\,, & q,\,\bar q \\
\frac{1}{2}\lambda_ab_0\,, & \gamma
\end{array}\right.\,,
\qquad 
M^2_a=
\left\{\begin{array}{ll}
\mu_5^2+m_a^2\,, & q,\,\bar q \\
0\,, & \gamma
\end{array}\right.
\,.
\gal
In the high-energy approximation:
\ball{j9}
E_a\approx |\b p_a|-\Lambda_a+\frac{M_a^2-\Lambda^2_a}{2|\b p_a|}\,.
\gal
The magnitude of the final momentum $k'$ can be determined by the energy conservation:
\ball{j11}
k'\approx k+\frac{1}{2}(\Lambda_3+\Lambda_4-\Lambda_1-\Lambda_2)+\frac{1}{4k}\left( M_1^2+M_2^2-M_3^2-M_4^2+\Lambda_3^2+\Lambda_4^2-\Lambda_1^2-\Lambda_2^2\right)\,.
\gal

Inspection of \eq{f55} and \eq{d12} reveals that the resonance emerges when 
\ball{j13}
q^2-M^2+2\Lambda|\b q|=0\,,
\gal
where $q$ is the momentum flowing through the propagator. The parameters $M$ and $\Lambda$, without a subscript, refer to the virtual state represented by the propagator. Denote $\cos\theta= \unit z\cdot \unit n$. In the $t$-channel, when $q=p_1-p_3$, the expression \eq{j13} vanishes when 
\ball{j15}
(\theta k)_0= -\Lambda\pm \sqrt{\Lambda^2-M^2-(\Lambda_1-\Lambda_3)(\Lambda_2-\Lambda_4)}\,,
\gal
whereas in the $u$-channel, when $q=p_1-p_4$, it vanishes when 
\ball{j17}
(\chi k)_0= -\Lambda\pm \sqrt{\Lambda^2-M^2-(\Lambda_1-\Lambda_4)(\Lambda_2-\Lambda_3)}\,,
\gal
where we introduced $\chi = \pi -\theta$. We took account of the fact that at high-energies, the scattering cross sections peak at either small $\theta$ or small $\chi$. Eqs.~\eq{j15} and \eq{j17} represent possible resonances in the $t$ and $u$ channels, respectively. A resonance occurs when there exists a set of helicities and polarizations such that  $(\theta k)_0$ or $(\chi k)_0$ are physical, meaning they are real and positive.

Consider two specific examples. 
\begin{description}

\item [Fermion-fermion scattering] $q(p_1)q(p_2)\to q(p_3) q(p_4)$. At high energies, the leading contribution to the cross section stems from the exchange of a virtual photon in the $t$-channel. Moreover, the fermion helicities do not change. As a result, \eq{j15} becomes:
\ball{j20}
(\theta k)_0
=
-\frac{\lambda b_0}{2}\pm \frac{|b_0|}{2}= |\lambda b_0|\delta_{\sgn(\lambda b_0),-1}\,.
\gal
A resonance emerges for the virtual photon's polarization  $\lambda =-\sgn(b_0)$ at $\sqrt{-t}=\theta k = b_0$. The impact of this resonance on the transport properties of chiral media was explored in \cite{Tuchin:2020gtz}. 

\item[Compton scattering] $q(p_1)\gamma(p_2)\to \gamma(p_3)q(p_4)$. In this case, the main contribution arises from the exchange of a virtual fermion  in the $u$-channel. In view of (approximate) helicity conservation \eq{j17} yields the following position of the resonance: 
\ball{j30}
(\chi k)_0= -\sigma\mu_5\pm \sqrt{-m^2}\,.
\gal
Since this expression is always complex, there is no resonance in the leading high-energy term.  

\end{description}

Apparently, only processes involving virtual photons  exhibit resonant behavior at the leading order. We verified numerically that this 
conclusion holds at any energy. 
At lower energies, spin-flip transitions play a more important role and, as a result, resonances exhibit stronger dependence on $\mu_5$.

We have not investigated whether the resonances can appear in virtual fermion states at higher orders of the perturbative expansion. However, in the next section we do demonstrate that both photon and fermion propagators can become resonant in a $2\to 3$ process at the leading order. 

\section{Bremsstrahlung}\label{sec:k}

\subsection{Resonances in fermion and photon propagators}

Consider now photon radiation in an external Coulomb field in chiral medium: $e(p_1)\to e(p_2)+\gamma(k)$. Denote the momentum transfer as $\b q$. According to \eq{f58}, the photon propagator has a familiar  infrared Coulomb pole at $\b q^2=0$. The other two poles correspond to zeros of denominators of fermion propagators $\tilde G(p_1-k)$ and $\tilde G(p_2+k)$. The denominator of the former propagator can be written as: 
\ball{k1}
\prod_{\sigma=\pm 1} \left[(p_1-k)^2-\mu_5^2-m^2-2\sigma\mu_5|\b p_1-\b k|\right]\,.
\gal
Using \eq{j5}, denoting by $\theta$ the angle between $\b p$ and $\b k$, and expanding in inverse powers of momentum, we obtain for each of the two factors in \eq{k1}:
\ball{k3}
(p_1-k)^2-\mu_5^2-m^2-2\mu_5|\b p_1-\b k|\approx -\theta^2 E\omega + 2(1-x)[\lambda b_0-\mu_5(\sigma+\sigma_1)]E
+\frac{b_0^2-2\lambda \mu_5\sigma_1 b_0 x-m^2x^2}{x}\,,
\gal
where $x=\omega /E$. At finite values of $\mu_5$ and $b_0$, we can neglect the last term, which is small. Clearly, this expression vanishes at the photon emission angle:
\ball{k5}
\theta_0= \sqrt{\frac{2(1-x)[\lambda b_0-\mu_5(\sigma+\sigma_1)]}{\omega}}\,,
\gal
provided that $\lambda b_0-\mu_5(\sigma+\sigma_1)>0$. In the case where $b_0$ and $\mu_5$ vanish, the last term in \eq{k3}, because of its negative sign, prevents \eq{k3} from vanishing at any $\theta$. 

A similar analysis applies to the other fermion propagator. Moreover, even the time-time component of the photon propagator $\tilde D^{00}(q)$, given by  \eq{f58}, that mediates between the fermion current and the Coulomb field, has a resonance. This is because the momentum transfer vanishes at a finite momentum transfer $\b q= \b p_2+\b k -\b p_1$. This can be seen by considering the smallest momentum transfer, which occurs when all three momenta are aligned. In this case:
\ball{k7}
q_\text{min}= |\b p_1|-|\b p_2|- |\b k| \approx E_1-\frac{m^2}{2E_1}+\sigma_1\mu_5 -E_2+\frac{m^2}{2E_2}-\sigma_2\mu_5
-\omega -\frac{\lambda b_0}{2}= \frac{m^2\omega}{2E_1E_2}+(\sigma_1-\sigma_2)\mu_5-\frac{1}{2}\lambda b_0\,.
\gal
Clearly, $q_\text{min}$ can be negative indicating that $|\b q|$ can vanish.

These conclusions are in agreement with the analysis of  \cite{Hansen:2023wzp}, which investigated bremsstrahlung at $\mu_5=0$. When an external field source possesses a magnetic moment (e.g.\ a heavy nucleus),  the incoming fermion couples to this magnetic moment  through the components $D^{ij}(q)$ of the photon propagator, as shown in \eq{f55}. An apparent resonance appears at $\b q^2=b_0^2$.  This anomalous contribution to bremsstrahlung was computed in \cite{Hansen:2022nbs} under the condition $\mu_5=0$.

\subsection{Regularization of  resonances}\label{sec:n}

The emergence of resonances indicates instability of the fermion and boson states in the chiral medium. Indeed this is precisely why the splitting processes discussed in \sec{sec:12} are possible at all. Since the fermion and boson states in the chiral medium are quasi-stationary they possess a finite width that is proportional to their total decay rate $\dot w$, which is given by \eq{h40} for photons and \eq{m11} for fermions.
The finite width $\dot w$ of the propagator screens the poles of both the photon and fermion propagators. However, in conductors, a more effective screening of the resonances of the photon propagator is achieved through conventional plasma polarization effects, which are characterized by the Debye mass at short distances.
 
 The resonances that occur in the fermion propagators, discussed in this section, reflect instability of fermion states in chiral matter with respect to spontaneous photon emission. The width of a quasi-static state is given by the total decay rate  as measured in the medium rest frame. Its inverse is the relaxation time $\tau=1/\dot w$. In the fermion rest frame the relaxation rate is $(E/m)\tau^{-1}$.  With the account of the finite width, the fermion propagators in \eq{k1} are regulated as follows:  
\ball{n1}
\prod_{\sigma=\pm} \left[(p_1-k)^2-\mu_5^2-m^2-2\sigma\mu_5|\b p_1-\b k|-iE_{\b p_1,\sigma_1}/\tau_{\b p_1,\sigma_1}\right]\,.
\gal
Due to the resonances, the bremsstrahlung total cross section is very sensitive  to $1\to 2$ rates. 

This regularization procedure can be employed to regularize photon and fermion propagators in other processes.

\section{Summary}\label{sec:s}

We considered an effective quantum electrodynamics in a chiral medium characterized by a constant chiral chemical potential $\mu_5$ and a chiral magnetic conductivity $b_0$. The chiral chemical potential signifies the chiral imbalance of the medium, whereas the chiral magnetic conductivity quantifies the strength of the electric  current induced by a magnetic field. The latter is a non-perturbative chiral magnetic effect, which we regard as a constitutive relation incorporated at the classical level of the effective theory. We employed this effective theory to analyze the scattering amplitude in a chiral medium and observed the emergence of resonances in certain chiral channels in $1\to 2$, $2\to 2$ and $2\to 3$ processes. 

We computed the paradigm $1\to 2$ processes, the chiral Cherenkov radiation and the pair production, that determine the width of quasi-stationary fermion and photon states in a chiral medium. Kinematic considerations require that $\mu_5$ enter the expressions for the resonances multiplied by the difference of incoming and outgoing quark helicities. Since at high energies, helicity flip is suppressed, the role of $\mu_5$ is also suppressed, while $b_0$ plays the leading role.  

Similar patterns are observed in processes involving a larger number of external particles. The widths of the resonances are determined by the $1\to 2$ decay rates, which in turn strongly influence the cross section of the resonant channels. A number of such processes were considered in the literature. In this work, we derived general conditions for the emergence of resonances and outlines the physical principles of their regularization. Generalization of our results to QCD is quite straightforward. 

We concentrated on chiraly isotropic systems with $\b b=0$, which have a simpler mathematical description. This by no means indicate that the results of this paper are restricted to  chiraly isotropic systems. For example, the chiral Cherenkov radiation is emitted at finite $\b b$ and shows a very similar resonant behavior. The primary distinction of an anisotropic chiral medium lies in the fact that the vector $\b b$ breaks the spherical symmetry and, as a result, the resonance structure depends on the relative angle between the photon direction and $\b b$ \cite{Lehnert:2004be,Huang:2018hgk,Tuchin:2018mte,Hansen:2024kvc,Tuchin:2025stl}.  

The instability of electrodynamics with the Chern-Simons term appearing in \eq{a1} is well-known  \cite{Carroll:1989vb}. This effect, referred to as the chiral magnetic instability, is triggered by the runaway modes $k<b_0$ of the electromagnetic field and is reflected in the pole at $k^2=b_0|\b k|$ in the photon propagator \eq{f60}. It  causes the transfer of chirality from the medium to the magnetic field through the inverse cascade process     \cite{Carroll:1989vb,Joyce:1997uy,Boyarsky:2011uy,Kharzeev:2013ffa,Khaidukov:2013sja,Kirilin:2013fqa,Akamatsu:2013pjd,Avdoshkin:2014gpa,Dvornikov:2014uza,Tuchin:2014iua,Manuel:2015zpa,Buividovich:2015jfa,Sigl:2015xva,Xia:2016any,Kaplan:2016drz,Kirilin:2017tdh,Tuchin:2018sqe,Mace:2019cqo}. While the chiral magnetic instability is indeed caused by the chiral magnetic current, it should not be confused with the instabilities in the scattering amplitude, which are the focus of this paper. The distinction between these instabilities becomes most apparent when $\b b$ is finite, while $b_0 = 0$. In this case, the scattering amplitudes are resonant, whereas the chiral magnetic instability is absent. 

In practical applications, the parameter $b_0$ may vary with time. A careful analysis performed in \cite{Tuchin:2025stl,Tuchin:2025bll,Tuchin:2026lka} reveals that when $b_0$ depends on time, the number of resonances can double depending on whether the asymptotic values of $b_0$  are finite or vanish.


It is noteworthy that this paper makes no mention of experimental measurements. This is because there are none, even though such measurements are technically possible. For instance, one can shoot an energetic electron through a Weyl semimetal to observe polarized THz radiation \cite{Hansen:2024kvc}, or measure the Brewster’s angle in topological insulators \cite{Stewart:2019xjh}. In relativistic heavy-ion collisions, one can study circularly polarized photons originating from the quark-gluon plasma \cite{Tuchin:2019jxd}. Similarly, cosmic rays may induce chiral Cherenkov radiation in the presence of a slowly varying axion field \cite{Huang:2018hgk}.

\acknowledgments
This work  was supported in part by the U.S. Department of Energy under Grant No.\ DE-SC0023692.

\bibliography{references}

\appendix
\section{Electromagnetic field at constant $b_0$}\label{sec:AppA}

\subsection{Energy of electromagnetic field}

Using \eq{f2} and \eq{f4}, we obtain: 
\ball{f8.1}
\frac{1}{2}\partial_t\int\left(\b E^2+ \b B^2\right)d^3x= \oint (\b B\times \b E)\cdot d\b a -b_0\int \b E\cdot \b B d^3x\,.
\gal
Noting that
\ball{8.3}
\partial_t\int \b A\cdot \b B d^3x= -2\int \b E\cdot \b B d^3x +\oint (\b A\times \b E)\cdot d\b a
\gal
we can rewrite the equation \eq{f8.1} as:
\ball{8.5}
\frac{1}{2}\partial_t\int\left(\b E^2+ \b B^2-b_0\b A\cdot \b B\right)d^3x=\oint (\b B\times \b E+\frac{1}{2}\b A\times \b E)\cdot d\b a\,. 
\gal
This equation represents the conservation of electromagnetic field energy, which is given by:
\ball{A1}
\mathcal{H}= \frac{1}{2}\int\left(\b E^2+ \b B^2-b_0\b A\cdot \b B\right)d^3x\,.
\gal
In view of \eq{f5}, $\mathcal{H}$  is gauge invariant. The complete energy-momentum tensor was obtained in \cite{Carroll:1989vb}.

To express the electromagnetic field energy as a Fock space operator, we substitute \eq{f27} into the left-hand side of \eq{8.5} and use the relations: 
\ball{A1}
\int \b A^*_{\b k',\lambda'}(x)\cdot \b A_{\b k,\lambda}(x)d^3x= 
\frac{(2\pi)^3}{2\omega_{\b k,\lambda}}\delta_{\lambda,\lambda'}\delta(\b k-\b k')
\gal
to derive:  
\ball{f30}
\mathcal{H}= \sum_\lambda \int \frac{d^3k}{(2\pi)^3} \omega_{\b k,\lambda}a^\dagger_{\b k,\lambda}a_{\b k,\lambda}\,,
\gal
up to an additive constant. The photon's energy is given by  \eq{f15}.

\subsection{Photon propagator}\label{sec:AppF}

The Feynman propagator is defined as the following time-ordered commutator:
\ball{f31}
D_F^{ij}(x)= D^{ij}(x)\theta(x^0)+D^{ij}(-x)\theta(-x^0)\,,
\gal
where 
\ball{f33}
D^{ij}(x)= \langle 0|A^i(x)A^j(0)|0\rangle=\sum_\lambda \int \frac{d^3p}{(2\pi)^3}\frac{\pi^{ij}_{\b p, \lambda}}{2\omega_{\b p, \lambda}}  e^{-ip\cdot x}\,.
\gal 
The product of the polarization vectors $\pi^{ij}_{\b p, \lambda}$ can be represented as a combination of symmetric and anti-symmetric matrices:
\ball{f35}
\pi^{ij}_{\b p, \lambda}= \epsilon^i_{\b p, \lambda}\epsilon^{j*}_{\b p, \lambda} = 
\frac{1}{2}\left( \delta^{ij}-\frac{p^ip^j}{\b p^2}\right)- \frac{i\lambda}{2}\epsilon^{ijk}\frac{p^k}{|\b p|}
\gal

Employing the integral representation of the step function
\ball{f37}
\theta(x^0)= -\frac{1}{2\pi i}\int_{-\infty}^\infty\frac{e^{-i s x^0}}{s+i\epsilon}ds
\gal
we can write 
\ball{f39}
D^{ij}(x)\theta(x^0)= i\sum_\lambda\int \frac{d^3k}{(2\pi)^3}\int_{-\infty}^\infty \frac{dk^0}{2\pi}\frac{\pi^{ij}_{\b k, \lambda}}{2\omega_{\b k, \lambda}}  e^{-ik\cdot x}\frac{1}{k^0-\omega_{\b k, \lambda}+i\epsilon}\,,
\gal
where the new integration variables are $\b k = \b p$ and $k^0=\omega_{\b k, \lambda}+s$. We can similarly represent
\ball{f41}
D^{ij}(-x)\theta(-x^0)=
i\sum_\lambda\int \frac{d^3k}{(2\pi)^3}\int_{-\infty}^\infty \frac{dk^0}{2\pi}\frac{\pi^{ij}_{-\b k, \lambda}}{2\omega_{-\b k, \lambda}}  e^{-ik\cdot x}\frac{1}{-k^0-\omega_{-\b k, \lambda}+i\epsilon}\,,
\gal
where now $\b k = -\b p$ and $k^0= -\omega_{-\b k, \lambda}-s$. 

It is evident from \eq{f11} and \eq{f13} that $\omega_{-\b k, \lambda}=\omega_{\b k, -\lambda}$. Also, \eq{f35} implies that $\pi^{ij}_{-\b k, \lambda}= \pi^{ij}_{\b k, -\lambda}$. Therefore, \eq{f31} now yields
\bal
D_F^{ij}(x)&= i\int \frac{d^3k}{(2\pi)^3}\int_{-\infty}^\infty\frac{dk^0}{2\pi} e^{-ik\cdot x}
\sum_\lambda
\left\{ \frac{\pi^{ij}_{\b k, \lambda}}{2\omega_{\b k, \lambda}} \frac{1}{k^0-\omega_{\b k, \lambda}+i\epsilon}
+ 
\frac{\pi^{ij}_{\b k, -\lambda}}{2\omega_{\b k, -\lambda}} \frac{1}{-k^0-\omega_{\b k, -\lambda}+i\epsilon}
\right\}\label{f51}\\
&=
i\int \frac{d^4k}{(2\pi)^4} e^{-ik\cdot x}
\sum_\lambda
 \frac{\pi^{ij}_{\b k, \lambda}}{(k^0)^2-\omega^2_{\b k, \lambda}+i\epsilon}\label{f32}\,.
\gal
Summing over $\lambda$ and using \eq{f15} and \eq{f35} we finally obtain for the propagator
\ball{f54}
D_F^{ij}(x)= \int \frac{d^4k}{(2\pi)^4} e^{-ik\cdot x} \tilde D_F^{ij}(k)\,,
\gal
where
\ball{f55}
\tilde D_F^{ij}(k)= \frac{i\left( \delta^{ij}-\hat k^i \hat k^j\right)k^2- b_0\epsilon^{ijr} k^r}{k^4-b_0^2\b k^2+i\epsilon}\,.
\gal
It is a  straightforward calculation to show that 
\ball{f57}
(-k^2\delta^{is}-ib_0 \epsilon^{irs}k^r)\tilde D^{sj}_F(k)= -i\left(\delta^{ij}-\hat k^i \hat k^j\right)\,,
\gal

In the Coulomb gauge $\b\nabla \cdot \b A=0$, while $A^0$ is finite. As a result, the space-space components of the propagator  $\tilde D_F^{\mu\nu}(k)$ are given by \eq{f55}, the time-time component $\tilde D_F^{00}$, which describes the instantaneous Coulomb interaction, reads:
\ball{f58}
\tilde D_F^{00}(k)=i/|\b k|^2\,,
\gal
 whereas the space-time components vanish $\tilde D_F^{0i}(k)=0$ \cite{Weinberg:1995mt}.  

The propagator in the Coulomb gauge, given by Eqs.~\eq{f55} and \eq{f58},  can be gauge-transformed into the boost-invariant expression \eq{f60} derived in \cite{Lehnert:2004hq}.
This is accomplished by the gauge transformation 
\ball{f62}
\tilde {\mathcal{D}}_F^{\mu\nu}(k)+ \zeta^\mu k^\nu + \zeta^\nu k^\mu\,,
\gal
with 
\ball{f63}
\zeta^0= \frac{ik^2 k^0}{2\b k^2(k^4-b_0^2\b k^2)}\,,\quad \zeta^i= -\frac{\zeta^0 k^i}{k^0}\,.
\gal

\section{Dirac field at constant $\mu_5$}\label{sec:AppE}

\subsection{Positive energy solutions of \eq{b1}}

The positive energy solution of \eq{b1} has a form:
\ball{b5}
\psi_{\b p,\sigma}(x)= e^{-ip\cdot x}u_{\b p, \sigma}=e^{-ip\cdot x}
\left(\begin{array}{c} \varphi _{\b p, \sigma} \\ \chi_{\b p, \sigma}\end{array}\right) \,,
\gal
where $p^\mu= (E_{\b p,\sigma},\b p)$, $\sigma$ is helicity, $\varphi$ and $\chi$ are left and right two-component spinors respectively. Substituting \eq{b5} into \eq{b1} and employing the chiral representation of $\gamma$-matrices we obtain:
\ball{b7}
\left(\begin{array}{cc}
-m & E_{\b p,\sigma}-\b p\cdot \b \sigma+\mu_5 \\
E_{\b p,\sigma}+\b p\cdot \b \sigma - \mu_5& -m\end{array}\right)\left(\begin{array}{c} \varphi _{\b p, \sigma} \\ \chi_{\b p, \sigma}\end{array}\right)=0\,.
\gal
Acting with the operator $\slashed p - \mu_5\gamma^5\gamma^0+m$ on this equation produces the set of equations:
\bal
\left[(E_{\b p,\sigma}-\b p\cdot \b \sigma + \mu_5)(E_{\b p,\sigma}+\b p\cdot \b \sigma - \mu_5)-m^2\right]\varphi _{\b p, \sigma}&=0\,,\label{b9}\\
\left[(E_{\b p,\sigma}+\b p\cdot \b \sigma - \mu_5)(E_{\b p,\sigma}-\b p\cdot \b \sigma + \mu_5)-m^2\right]\chi _{\b p, \sigma}&=0\,. \label{b10}
\gal
These equations are diagonalized only if the functions $\varphi_{\b p, \sigma}$ and $\chi_{\b p, \sigma}$ are proportional to the  helicity eigenstates $\xi_{\b p, \sigma}$ that obey the equation:
\ball{b12}
\unit p \cdot \b\sigma \xi_{\b p, \sigma}=\sigma \xi_{\b p, \sigma}\,,
\gal
and normalized as $\xi_{\b p, \sigma}^\dagger \xi_{\b p, \sigma'}=\delta_{\sigma,\sigma'}$, where $\sigma, \sigma'=\pm 1$. Suppose that $\varphi_{\b p, \sigma}= N_\sigma\xi_{\b p, \sigma}$, where $N_\sigma$ is a normalization constant. It follows from \eq{b7} that 
\ball{b14}
\chi_{\b p, \sigma}= \frac{N_\sigma}{m}(E_{\b p,\sigma}+\sigma|\b p|-\mu_5)\xi_{\b p, \sigma}\,.
\gal
Eq.~\eq{b9} then implies:
\ball{b16}
E_{\b p,\sigma}= \sqrt{(\sigma|\b p|-\mu_5)^2+m^2}\,.
\gal

The Dirac spinors are normalized as follows:
\ball{b18}
u_{\b p, \sigma}^\dagger u_{\b p, \sigma}=2E_{\b p,\sigma}\,.
\gal
The left-hand side of \eq{b18} can be expressed as  
\ball{b20}
u_{\b p, \sigma}^\dagger u_{\b p, \sigma}= \varphi_{\b p, \sigma}^\dagger \varphi_{\b p, \sigma}+\chi_{\b p, \sigma}^\dagger \chi_{\b p, \sigma}= \frac{|N_\sigma|^2}{m^2} \left[ (E_{\b p,\sigma}+\sigma|\b p|-\mu_5)^2+m^2\right]
=\frac{|N_\sigma|^2}{m^2} 2E_{\b p, \sigma}\left( E_{\b p, \sigma}+\sigma|\b p|-\mu_5\right)\,,
\gal  
which implies that:
\ball{b22}
|N_\sigma| = \frac{m}{\sqrt{E_{\b p, \sigma}+\sigma|\b p|-\mu_5}}= \sqrt{E_{\b p, \sigma}-\sigma|\b p|+\mu_5}\,.
\gal
Finally, the particle spinors are given by \eq{b24}.

\subsection{Negative energy solutions}

We seek the negative energy solutions to \eq{b1} in the form:
\ball{c5}
\psi_{\b p,\sigma}(x)= e^{ip\cdot x}v_{\b p, \sigma}=e^{ip\cdot x}
\left(\begin{array}{c} \varphi' _{\b p, \sigma} \\ \chi'_{\b p, \sigma}\end{array}\right) \,,
\gal
where $p^\mu= (E'_{\b p,\sigma},\b p)$. Substitution into \eq{b1} now yields:
\ball{c7}
\left(\begin{array}{cc}
-m & -E'_{\b p,\sigma}+\b p\cdot \b \sigma+\mu_5 \\
-E'_{\b p,\sigma}-\b p\cdot \b \sigma - \mu_5& -m\end{array}\right)\left(\begin{array}{c} \varphi' _{\b p, \sigma} \\ \chi'_{\b p, \sigma}\end{array}\right)=0\,.
\gal
The solution in terms of the normalized helicity eigenstates is:
\bal
\varphi'_{\b p, \sigma}&= N'_\sigma\xi_{\b p, \sigma}\,,\label{c9}\\
\chi'_{\b p, \sigma}&= -\frac{N'_\sigma}{m}(E'_{\b p,\sigma}+\sigma|\b p|+\mu_5)\xi_{\b p, \sigma}\,,\label{c10}
\gal
with the dispersion relation: 
\ball{c12}
E'_{\b p,\sigma}=E_{\b p,-\sigma} =\sqrt{(\sigma|\b p|+\mu_5)^2+m^2}\,.
\gal
The normalization constant is computed as in \eq{b18}-\eq{b22} with the result:
\ball{c14}
|N'_\sigma|= \frac{m}{\sqrt{E'_{\b p, \sigma}+\sigma|\b p|+\mu_5}}=\sqrt{E'_{\b p, \sigma}-\sigma|\b p|-\mu_5}\,.
\gal
This yield the final expression for the anti-particle spinors given by \eq{c24}.


\subsection{Gordon's identity}


Consider the Dirac equation at finite chiral chemical potential \eq{b1} for the positive energy solutions:
\ball{e1}
\left( p_\mu \gamma^\mu -\gamma^5\gamma^\mu a_\mu-m\right)u_{\b p,\sigma}=0\,.
\gal
The adjoint spinor obeys the equation
\ball{e2}
\bar u _{\b p,\sigma}\left( p_\mu \gamma^\mu -\gamma^5\gamma^\mu a_\mu-m\right)=0\,.
\gal
It can be shown that the spinors satisfy the Gordon's identity:
\ball{e5}
\bar u _{\b p',\sigma'}\gamma^\mu u_{\b p,\sigma} 
= \frac{1}{2m}\bar u _{\b p',\sigma'}\left\{ p^\mu+{p'}^\mu+ i\sigma^{\mu\nu}({p'}_\nu-p_\nu) 
-2i \sigma^{\mu\nu} \gamma^5 a_\nu\right\} u_{\b p,\sigma}\,.
\gal
It is straightforward to verify that 
\bal
\bar u _{\b p',\sigma'}(\slashed p-\slashed p') u_{\b p,\sigma} =0\,,
\gal
which is consistent with the charge conservation requirement.   


Using the Gordon identity, the current density is:
\ball{e14}
j^\mu = \frac{1}{2E_{\b p, \sigma}}\bar u _{\b p,\sigma}\gamma^\mu u_{\b p,\sigma} = 
 \frac{1}{2E_{\b p, \sigma}}\frac{1}{2m}\bar u _{\b p,\sigma}\left( 2p^\mu
-2i \sigma^{\mu\nu} \gamma^5 a_\nu\right) u_{\b p,\sigma} 
= \frac{1}{E_{\b p, \sigma}}\left(E_{\b p, \sigma},\b p -\mu_5 \langle \b \sigma \rangle \right)\,.
\gal
where $\langle \b \sigma \rangle =\xi^\dagger_{\b p,\sigma} \b \sigma\xi_{\b p,\sigma}$. 

Given the momentum $\b p$ with the polar and azimuthal angles $\theta$ and $\phi$, the explicit helicity eigenstates read
\ball{e16}
\xi_{\b p,+}= \left(\begin{array}{c} \cos\frac{\theta}{2}  \\ e^{i\phi}\sin\frac{\theta}{2}  \end{array}\right)\,,\qquad
\xi_{\b p,-}= \left(\begin{array}{c} -e^{-i\phi}\sin\frac{\theta}{2}  \\ \cos\frac{\theta}{2}  \end{array}\right)\,.
\gal
Using these we compute: 
\ball{e18}
\langle \b \sigma \rangle= \sigma \frac{\b p}{|\b p|}\,.
\gal

It is remarkable that even a particle at rest have non-vanishing current 
\ball{e20}
\lim_{\b p\to 0}\b j = -\frac{\mu_5 \sigma \unit p  }{\sqrt{m^2+\mu_5^2}}\,,
\gal
which is a reflection of non-zero Berry curvature \cite{Son:2012zy}.


\subsection{Fermion propagator}

The fermionic Green function $G(x)$ obeys the equation: 
\ball{d3}
\left(i\slashed \partial - \gamma^5\gamma^0\mu_5-m\right)G(x)= i \delta(x)\,.
\gal
In the momentum space \eq{d3} reads:
\ball{d5}
\left(\slashed p -  \gamma^5\gamma^0\mu_5-m\right)\tilde G(p)= i 
\gal
Using the properties of $\gamma$ matrices, we compute: 
\ball{d7}
\left(\slashed p -  \gamma^5\gamma^0\mu_5 +m\right)\left(\slashed p -  \gamma^5\gamma^0\mu_5-m\right)
=p^2-\mu_5^2-m^2+2\mu_5\gamma^5\gamma^0 \b \gamma\cdot \b p
\gal
and 
\ball{d9}
\left(p^2-\mu_5^2-m^2-2\mu_5\gamma^5\gamma^0\b \gamma\cdot \b p\right)\left(p^2-\mu_5^2-m^2+2\mu_5\gamma^5\gamma^0\b \gamma\cdot \b p\right)
=\left(p^2-\mu_5^2-m^2\right)^2-4\mu_5^2\b p^2\,.
\gal
Using the fact that in the chiral representation $\gamma^5\gamma^0\b\gamma=\b \Sigma$, allows us to cast the solution to \eq{d5} in the form given by \eq{d12}.
It is worth noting that the two matrices enclosed within the two pairs of parentheses in the numerator of \eq{d12} commute.

\end{document}